\documentclass[floatfix,twocolumn,nofootinbib,superscriptaddress,a4paper,preprintnumbers]{revtex4-1}

\usepackage{amsmath}
\usepackage{amssymb}
\usepackage{graphicx}
\usepackage[T1]{fontenc}
\usepackage{slashed}
\usepackage[utf8]{inputenc}
\usepackage{xspace}
\usepackage{color}
\usepackage{ulem}

\newcommand{\fb}{\textrm{fb}}

\newcommand{\calO}{{\cal O}}

\newcommand{\AddrPeking}{Center for High-Energy
Physics, Peking University, Beijing, 100871, P. R. China}
\newcommand{\AddrKavli}{State Key Laboratory of Theoretical Physics
and Kavli Institute for Theoretical Physics, China (KITPC),
Institute of Theoretical Physics, Chinese Academy of Sciences,
Beijing 100190, P. R. China}
\newcommand{\AddrChengdu}{School of Physical Electronics,
University of Electronic Science and Technology of China,
Chengdu 610054, P. R. China}
\newcommand{\AddrBeijing}{Institute of Physics Chinese Academy of sciences, Beijing 100190, P. R. China}

\begin{document}

\title{Interpreting $750$ GeV Diphoton Excess with R-parity Violating Supersymmetry}

\author{Ran Ding}
\affiliation{\AddrPeking}

\author{Li Huang}
\affiliation{\AddrKavli}

\author{Tianjun Li}
\affiliation{\AddrKavli}
\affiliation{\AddrChengdu}

\author{Bin Zhu}
\affiliation{\AddrBeijing}

\begin{abstract}
We propose an supersymmetric explanation of the diphoton excess in the Minimal Supersymmetric Standard Model with the leptonic R-parity violation. Where the sneutrino serves as the 750 GeV resonance and produced through quark anti-quark annihilation. With introducing appropriate trilinear soft parameters, we show that the diphoton branching ratio is significantly enhanced compared with the conventional MSSM. The parameter space favored by diphoton excess strongly indicates the mass of smuon and stau fall into the range $375-410$ (375-480) GeV, which depending on the electroweakino masses. In addition, the $R$-parity-violating trilinear couplings involved with second generation quarks are both favored by compatibility of diphoton excess and low-energy constraints.

\end{abstract}

\maketitle


{\bf 1.~Introduction}--Recently, both ATLAS and CMS reported an excess on diphoton channel around the invariant mass $M\simeq750$ GeV in the run II data of LHC at $\sqrt{s}=13$ TeV. For ATLAS observation, the local significance of the excess reaches $3.9~\sigma$ with a best-fit width of about 45 GeV ($\Gamma/M\simeq0.06$)~\cite{bib:ATLAS_diphoton}. While for CMS observation, the local significance is $2.6~\sigma$ and the best-fit prefers a narrow width~\cite{bib:CMS_diphoton}. The corresponding signal cross sections can be estimated as:
\begin{equation}
\sigma^{\rm 13 TeV}_{pp\to\gamma\gamma} \simeq
\left\{\begin{array}{lll}
(10\pm3)~\fb~&\hbox{for ATLAS~\cite{bib:ATLAS_diphoton}},\\
(6\pm3)~\fb~&\hbox{for CMS~\cite{bib:CMS_diphoton}}.
\end{array}\right.
\label{eq:sigma813}
\end{equation}

Although it may be eventually identified as a fluctuation, the possibility of a new resonance is likely to be strong hints for new physics beyond the standard model (BSM). Which has inspired many model building efforts in various phenomenological frameworks~\cite{Franceschini:2015kwy,Pilaftsis:2015ycr,Buttazzo:2015txu,Knapen:2015dap,Angelescu:2015uiz,Nakai:2015ptz,
Backovic:2015fnp,Mambrini:2015wyu,Harigaya:2015ezk,Molinaro:2015cwg,Petersson:2015mkr,Gupta:2015zzs,
Bellazzini:2015nxw,Low:2015qep,Ellis:2015oso,McDermott:2015sck,Higaki:2015jag,Bai:2015nbs,Aloni:2015mxa,
Falkowski:2015swt,Csaki:2015vek,Agrawal:2015dbf,Ahmed:2015uqt,Chakrabortty:2015hff,Bian:2015kjt,Curtin:2015jcv,
Fichet:2015vvy,Chao:2015ttq,Demidov:2015zqn,No:2015bsn,Becirevic:2015fmu,Cox:2015ckc,Kobakhidze:2015ldh,
Matsuzaki:2015che,Dutta:2015wqh}.

One of the most intriguing problems is then whether or not this excess can be interpreted in the framework of Supersymmetry (SUSY). In this direction, current works focus on the heavy Higgs candidates in the Minimal Supersymmetric Standard Model (MSSM) and the Next-to-Minimal Supersymmetric Standard Model (NMSSM)~\cite{Angelescu:2015uiz}. Unfortunately, it is found that the diphoton branching ratio reaches only $\calO(10^{-6})$ even in the case of $\tan\beta\sim 1$ which is the lower limit required by the Renormalization Group Equation (RGE) of Yukawa couplings. As a result, one has to introduce extra vector-like fermions to alleviate the discrepancy of diphoton rate between the model prediction and experimental requirement. In this paper, we suggest a novel approach to explain such excess without the need of any ad-hoc addition of extra particles. We consider the framework of Leptonic $R$-parity-violating (LRPV) MSSM, where the 750 GeV resonances is identified with the sneutrino and produced via quark anti-quark annihilation. The diphoton excess is then originated from its loop-induced decay. With introducing appropriate LRPV soft breaking trilinear terms, the diphoton branching ratio receives significant enhancement compared with conventional MSSM due to the contribution of sleptons in the loop.

The organization of paper is as follows: In section 2, we introduce our model and illustrate the mechanism of diphoton enchantment. In section 3, we then explore the parameter space in our model with taking into account relevant LHC limits and low-energy constraints. The last section is devoted to conclusion.

{\bf 2.~LRPV MSSM}-- We start with the superpotential of LRPV model~\cite{Barbier:2004ez}:
\begin{eqnarray}
W &=& Y_d^{ij}Q_iH_dD^c_j+ Y_u^{ij}Q_iH_uU^c_j +Y_e^{ij}L_iH_dE^c_j+\mu H_uH_d\nonumber\\
  &+&\frac{1}{2}\lambda_1^{ijk}L_iL_jE^c_k+\lambda_2^{ijk}L_iQ_jD^c_k~.~\,
   \label{eq:st}
\end{eqnarray}
In above equation, the $SU(2)_L$ and $SU(3)_C$ indices have been suppressed. $i,~j,~k=1,~2,~3$ are the the family indices and a summation is implied. $Q_i$ ($L_i$) are the $SU(2)_L$ doublet quark (lepton) superfields. $D^c_j$ and $U^c_j$ ($E^c_j$) are the $SU(2)_L$ singlet down- and up-quark (electron) superfields, respectively. $\lambda_1$ and $\lambda_2$ are trilinear couplings. The lepton number is automatically violated by $\lambda_1^{ijk}L_iL_jE^c_k$ and $\lambda_2^{ijk}L_iQ_jD^c_k$ operators and the first term is anti-symmetric in $i,j$ indices. In our model, sneutrino serves as $750$ GeV resonance. We therefore only list the relevant soft interactions for sneutrino sector:
\begin{eqnarray}
- L_{\rm soft} &=&T_{\lambda_1}^{ijk}\tilde{\nu}_{iL} \tilde{e}_{jL} \tilde{e}^*_{kR} +T_{\lambda_2}^{i j k} \tilde{\nu}_{iL}\tilde{d}^\beta_{jL} \tilde{d}^{*\gamma}_{kR}\delta_{\beta \gamma}\nonumber\\
&+&\tilde{\nu}^*_{iL} (m_{\tilde{l}}^{2})^{ij} \tilde{\nu}_{jL}~,~\,
\label{eq:sf}
\end{eqnarray}
where the fields with tilde denote the scalar fermion superpartners. As is shown in Eq.~\ref{eq:st} and~\ref{eq:sf}, the production and decay properties of sneutrino are determined by $\lambda_1^{ijk}$, $\lambda_2^{ijk}$, $T_{\lambda_1}^{ijk}$ and $T_{\lambda_2}^{ijk}$. Here we only consider the first generation sneutrino $\tilde{\nu}_e$ as a illustration. It is straightforward to include the other generations by multiplying a factor to the signal strength if one takes the degenerate mass spectrum. In order to compute the loop-induced decay $\tilde{\nu}_e\rightarrow\gamma\gamma$, we divide the sneutrino into its CP-even and CP-odd part,
 \begin{align}
 \tilde{\nu}_e=\frac{1}{\sqrt{2}}(\tilde{\nu}_e^{+}+i\tilde{\nu}_e^{-}).
 \end{align}
We first investigate the contributions from fermionic loop. This
calculation is analogous to the CP-even/odd neutral Higgs decay into diphoton and into di-gluon in the MSSM. One obtains following partial widths~\cite{BarShalom:1998xz},
 \begin{eqnarray}
 \Gamma(\tilde\nu_e^{\pm}\rightarrow\gamma\gamma)&=&\frac{\alpha^2m_{\tilde\nu_e^{\pm}}^3}{512\pi^3}
 \left|\sum_{j=1}^{3}\frac{N_c}{m_{d_j}}e_{d_j}^2\lambda_{2}^{1jj}A_{1/2}^{\pm}(\tau_{d_j})\right. \nonumber \\
 &+& \left. \sum_{j=2}^{3}\frac{1}{m_{l_j}}\lambda_{1}^{1jj}A_{1/2}^{\pm}(\tau_{l_j})\right|^2,
\label{eq:ga}
\end{eqnarray}

\begin{eqnarray}
\Gamma(\tilde\nu_e^{\pm}\rightarrow gg)&=&\frac{\alpha_s^2m_{\tilde\nu_e^{\pm}}^3}{256\pi^3}
 \left|\sum_{j=1}^{3}\frac{1}{m_{d_j}}\lambda_{2}^{1jj}A_{1/2}^{\pm}(\tau_{d_j})\right|^2,
 \label{eq:gg}
 \end{eqnarray}
where the sum in Eq.~\ref{eq:ga} (Eq.~\ref{eq:gg}) runs over all down- fermions (down–quarks). Furthermore, $N_c=3$ is the number of colors and $\alpha$ ($\alpha_s$) is the QED (QCD) fine structure constant. $e_{d_j}=-1/3$ is the electric charge of down-type quarks, $\tau=4 m^2/m_{\tilde\nu}^2$ with $m$ the mass of particles running in loop. The loop function $A_{1/2}^{\pm}$ is given by~\cite{Gunion:1989we},
 \begin{align}
 A_{1/2}^{+}=&-2\tau(1+(1-\tau)f(\tau)),\nonumber\\
 A_{1/2}^{-}=&-2\tau f(\tau),\nonumber\\
 f(\tau)=&
	\begin{cases}
		\arcsin^2[1/\sqrt{\tau}],  & \mbox{if } \tau \geq 1 \\
		-\dfrac{1}{4}[\ln \dfrac{1+\sqrt{1-\tau}}{1-\sqrt{1-\tau}} - i \pi]^2, & \mbox{if } \tau < 1.
    \end{cases}
 \end{align}
The structure of $A_{1/2}^{\pm}$ implies that $A_{1/2}^{\pm}/m\rightarrow0$ when $m\rightarrow0$. This feature indicates that
loop contribution from quarks and leptons are highly suppressed except for the third generation, i.e., bottom quark and tau lepton. While including the contribution of third generation fermions also introduce a new decay modes like $\Gamma(\tilde\nu\rightarrow b \bar{b})$ and $\Gamma(\tilde\nu\rightarrow \tau^{+}\tau^{-})$, which considerably suppress the branching ratio of diphoton mode. Moreover, the production of sneutrino through $b\bar{b}$ channel is expected to be much smaller than that of first two generation quarks. Due to the above reasons, we do not
consider the case of third generation and assuming $j,k=1,2$ for $\lambda_{1}^{1jk}$ and $\lambda_{2}^{1jk}$.

We then take into account contribution from sfermions. As we will show later, which plays a crucial role in diphoton enhancement. The decay modes in Eq.~\ref{eq:ga} and \ref{eq:gg} can be safely neglected in this case. Notice that we can simply treated CP-even $\tilde\nu^{+}$ as the resonance since the sfermion loops can not give any contribution to CP-odd $\tilde\nu^{-}$. The corresponding partial widths are given by,
\begin{eqnarray}
 \delta\Gamma(\tilde\nu_e^{+}\rightarrow\gamma\gamma)&=&\frac{\alpha^2m_{\tilde\nu_e^{\pm}}^3}{512\pi^3}
 \left|\sum_{j=1}^{3}\frac{N_c}{2m_{\tilde{d}_j}^2}e^2_{d_j}\sin2\theta_2 T_{\lambda_2}^{1jj}A_{0}^{+}(\tau_{\tilde{d}_j})\right.\nonumber\\
 &+& \sum_{j=2}^{3}\left.\frac{1}{2m_{\tilde{l}_j}^2} T_{\lambda_1}^{1jj}\sin2\theta_1 A_{0}^{+}(\tau_{\tilde{l}_j})\right|^2,
 \label{eq:ga2}
\end{eqnarray}

\begin{eqnarray}
 \delta\Gamma(\tilde\nu_e^{+}\rightarrow gg)&=&\frac{\alpha_s^2m_{\tilde\nu_e^{\pm}}^3}{256\pi^3}
 \left|\sum_{j=1}^{3}\frac{1}{2m_{\tilde{d}_j}^2}T_{\lambda_2}^{1jj}\sin2\theta_2 A_{0}^{+}(\tau_{\tilde{d}_j})\right|^2,
 \label{eq:ga3}
\end{eqnarray}
where
\begin{align}
A_{0}^{+}=\tau(1-\tau f(\tau)).
\end{align}
In this case, the mass eigenstates of sleptons (only second and third generations are involved since $T_{\lambda_2}^{1jj}$ is anti-symmetric in $1,j$ indices.) and squarks (can be both three generations) do not coincide with the gauge eigenstates due to introduce non-zero mixing trilinear terms. Here $\theta_1/\theta_2$ being the mixing angle between left-handed and right-handed slepton/squarks, respectively. In order to enhance diphoton partial width as much as possible, we assume maximal mixing scenario in this paper, i.e., setting $\theta_1=\theta_2=\pi/4$. As a consequence, the mass eigenstates tend to have a large splitting thus the contributions from heavier ones can be ignored in the loop calculation.

One immediately find that both $\delta\Gamma(\tilde\nu^{+}\rightarrow\gamma\gamma)$ and $\delta\Gamma(\tilde\nu^{+}\rightarrow gg)$
are proportional to $T_\lambda/m_{\tilde{f}}$. To enhance the diphoton partial width effectively, one thus prefers large soft trilinear term $T_\lambda$ and relatively light sfermion. Unfortunately, the di-gluon partial width is also received enhancement in this way and the ratio of two modes is roughly estimated as $\alpha_s^2/\alpha^2$. As a result, the total effect of sfermion loops actually reduces the diphoton branching ratio. The simplest way to solve this dilemma is choose $T_{\lambda_2}^{ijk}=0$, which forbidding the contribution of squarks and only leaving sleptons in the loop. One thus has
\begin{eqnarray}
\Gamma_{\gamma\gamma} &\equiv& \delta\Gamma(\tilde\nu_e^{+}\rightarrow\gamma\gamma)\nonumber\\
&=&\frac{\alpha^2m_{\tilde\nu_e^{\pm}}^3}{512\pi^3}
 \left|\sum_{j=2}^{3}\frac{1}{2m_{\tilde{l}_j}^2} T_{\lambda_1}^{1jj}\sin2\theta_1 A_{0}^{+}(\tau_{\tilde{l}_j})\right|^2,
\label{eq:ga2}
\end{eqnarray}
and $\delta\Gamma(\tilde\nu_e^{+}\rightarrow g g)=0$.

We finally obtain a valid recipe in LRPV MSSM which successfully enhances diphoton partial width while avoiding unwanted di-gluon mode.

In the rest of this section, we deal with other important decay modes of sneutrino:
\begin{itemize}
\item  Decaying into first two generation di-quarks via $LQD^c$ operator with the partial width,
    \begin{align}
    \Gamma_{d\bar{d}}\equiv\Gamma(\tilde{\nu}_e^{+}\rightarrow d_j \bar{d}_k )=\frac{3}{16\pi}|\lambda_2^{1jk}|^2m_{\tilde{\nu}_e^+}.
    \label{eq:dj}
    \end{align}
It is the inverse process of sneutrino production thus unavoidable and  dominates the branching ratios in general. This decay mode leads to dijet final states and impose stringent constraint on $\lambda_2^{1jk}$.

\item Decaying into di-lepton via $LLE^c$ operator, which is severely limited by LHC di-lepton resonance searches~\cite{Aad:2014cka}. This decay mode can not be forbidden kinematically due to the lightness of leptons. We thus assuming $\lambda_1^{ijk}$ is negligible to remove this dangerous mode.

\item Decaying into squark and slepton pairs due to the soft trilinear terms $T_{\lambda_2}\tilde{\nu}_L\tilde{d}_L\tilde{d}^*_R$ and $T_{\lambda_1}\tilde{\nu}_L\tilde{e}_L\tilde{e}^*_R$, respectively. Among them, the di-squark mode is automatically vanished since we have taken $T_{\lambda_2}=0$. While for di-slepton mode, we must keep large $T_{\lambda_1}$ since it is also responsible for enhancing the diphoton partial width. We therefore forbid di-slepton mode kinematically by setting $m_{\tilde\mu,\tilde\tau}>375$ GeV.

\item Decaying into electroweakinos (neutrlinos and charginos) through gauge interaction. They have following partial widths~\cite{Barger:1989rk}:
\begin{eqnarray}
\Gamma_{\chi l}&\equiv&\Gamma(\tilde{\nu}_e^{+}\rightarrow\tilde{\chi}_a^0\nu_e,\tilde{\chi}_a^{+}e^-)\nonumber\\
&=&\frac{C g^2}{16\pi}m_{\tilde{\nu}_e^{+}}\left(1-\frac{m_{\tilde{\chi}_a^{+}}^2}{m_{\tilde{\nu}_e^{+}}^2}\right)^2.
\label{eq:chi}
\end{eqnarray}
Where the coefficient $C=|N_{a2}|^2$ ($C=|V_{a1}|^2$) for the neutralinos (charginos) case. with $N_{a2}$ and $V_{a1}$ the elements of mixing matrix.

\item Finally, decaying into neutral gauge bosons $Z\gamma$ and $ZZ$ via the same loop diagram as the diphoton mode. However, their contributions to the total decay width are much smaller than other modes discussed above. We therefore neglect these two modes in the following discussion and numerical calculations.
\end{itemize}
In summary, one leaves three modes in the decay pattern of sneutrino:
$\tilde\nu_e\rightarrow\gamma\gamma$ via slepton loops, $\tilde{\nu}_e\rightarrow d_j \bar{d}_k$ and $\tilde{\nu}_e\rightarrow\tilde{\chi}^0\nu_e,\tilde{\chi}^{+}e^-$. The total decay width is then $\Gamma_{\rm tot}=\Gamma_{\gamma\gamma}+\Gamma_{d\bar{d}}+\Gamma_{\chi l}$, with the branching ratio of diphoton and di-quark modes are respectively ${\rm BR}_{\gamma\gamma}=\Gamma_{\gamma\gamma}/\Gamma_{\rm tot}$ and ${\rm BR}_{d\bar{d}}=\Gamma_{d\bar{d}}/\Gamma_{\rm tot}$.

{\bf 3.~Diphoton excess and LHC constraints}--In this section, we investigate the diphoton signal rate in our model. For this purpose, we use {\tt SARAH}~\cite{sarah} to generate {\tt UFO} model file~\cite{Degrande:2011ua}, and {\tt MadGraph5\_aMC@NLO}~\cite{MG5} to calculate the production cross section of sneutrino with {\tt CTEQ6L1}~\cite{Nadolsky:2008zw} parton distribution function (PDF). Based on the discussion in section 2, we have following input parameters:
[$m_{\tilde{l}},~m_{{\tilde\chi}^\pm},m_{{\tilde\chi}^0},~T^{1jj}_{\lambda_1}, ~\lambda_2^{1jk}$]. Where the independent components of $T^{1jj}_{\lambda_1}$ and  $\lambda_2^{1jk}$ are $T^{1jj}_{\lambda_1}=T^{122,133}_{\lambda_1}$ and $\lambda_2^{1jk}=\lambda_2^{111,112,121,122}$, respectively.

Before showing our results, we make some comments on the choice of parameters in the numerical analysis. We work in the framework of simplified SUSY models, with only involving sleptons and electroweakinos in the mass spectrum. In this scenario, one can take either light or heavy electroweakinos, which has significant impact on the total decay width of sneutrino. The current LHC limits on RPV SUSY models are reviewed in Ref.~\cite{Redelbach:2015meu}. Those related to our model are constraints on $LLE^c$ and $LQD^c$ interactions. For $LLE^c$ operator, by assuming $\tilde{\chi}^{0}_1$ as a lightest supersymmetric particle (LSP) and using simplified models, one obtains following approximate upper bounds on superpartner masses:
\begin{itemize}
  \item gluino masses $m_{\tilde g} > 950$ GeV,
  \item light stop masses $m_{{\tilde t}_1} > 820$ GeV,
  \item charged slepton masses $m_{\tilde l} > 240$ GeV,
  \item sneutrino masses $m_{\tilde\nu} > 400$ GeV,
  \item wino-like chargino masses $m_{\tilde{\chi}^\pm_1} > 450$ GeV.
\end{itemize}
It is obviously that above limits can be easily escaped through chosing heavy gluino and squarks and arranging the mixing matrix of charginos. Moreover, noticed that we assumed the relevant couplings $\lambda_1$ is negligible in our model. Therefore, these bounds are still relaxed significantly even we do not apply the heavy spectrum. Similarly, constraints based on $LQD^c$ operators also do not threat our model since the relevant searches are mainly investigated with stop-pair production and setting stop masses up to 1 TeV. Based on above discussion,
we choose $m_{{\tilde\chi}^{\pm},\tilde{\chi}^{0}}=350$ (800) GeV for the case of light (heavy) electroweakinos, and taking gluino and squarks are heavier than 1 TeV. In order to extract the key features from the parameter space, we further make following assumption:
\begin{itemize}
  \item {The soft trilinear couplings $T_{\lambda_1}$ are fixed as $T^{122}_{\lambda_1}=T^{133}_{\lambda_1}=10$ TeV.}
  \item {Due to the PDF dependence, the luminosity ratios between 8 TeV and 13 TeV are distinct for first and second generation quarks. It is then interesting to examine their contributions separately. For this aim, We treated $\lambda_2^{111}$ and $\lambda_2^{122}$ as independent operators. Furthermore, the non-diagonal couplings $\lambda_2^{112}$ and
  and $\lambda_2^{121}$ are combined as a single operator (labeled by $\lambda_2^{12,21}$) with $\lambda_2^{112}=\lambda_2^{121}$ to account for mixed contributions of first two generation quarks. In the numerical calculation, we assume that each operators contributes one at a time. Finally , we use $\lambda^{\rm tot}_2$ to denote the universal couplings with $\lambda_2^{111}=\lambda_2^{122}=\lambda_2^{112}=\lambda_2^{121}$.}
\end{itemize}

In table~\ref{tab:pro_snu}, we list cross sections of sneutrino production at LHC $13$ TeV ($8$ TeV) respectively correspond to couplings $\lambda_2^{111}$, $\lambda_2^{122}$, $\lambda_2^{112,21}$ and $\lambda_2^{\rm tot}$. Where the typical value is taken as $|\lambda_2|=1$. Their cross sections fall into the region $\calO(10-100)$ pb and possessing following order,
\begin{align}
\sigma^{13 \rm TeV}_{pp\to\tilde{\nu}_e}(\lambda_2^{111})>\sigma^{13 \rm TeV}_{pp\to\tilde{\nu}_e}(\lambda_2^{112,21})
>\sigma^{13 \rm TeV}_{pp\to\tilde{\nu}_e}(\lambda_2^{122}).
\label{eq:pro}
\end{align}
Which is resulted from PDF dependence of first and second generation quarks. In the narrow width approximation, the diphoton signal rate can be calculated as
\begin{align}
\sigma^{13 \rm TeV}_{pp\to \gamma\gamma}\simeq\sigma^{13 \rm TeV}_{pp\to\tilde{\nu}_e}|\lambda_2^{1jk}|^2\cdot{\rm BR}_{\gamma\gamma}.
\label{eq:diphoton13}
\end{align}

Currently, the most stringent LHC constraint for our model coming from dijet and diphoton resonance searches performed at LHC 8 TeV. Corresponding upper bounds at $95\%$ confidence level yield $\sigma^{8 \rm TeV}_{pp\to jj}<2.5$ pb~\cite{CMS:2015neg} and $\sigma^{8 \rm TeV}_{pp\to \gamma\gamma}<1.5$ fb~\cite{Aad:2015mna,CMS:2014onr}, respectively. Since they have different impact on the model parameter space, we discuss them one by one. We start with dijet constraint. The cross section of dijet final states resulted from $\tilde{\nu}_e$ decay is given by,
\begin{align}
\sigma^{8 \rm TeV}_{pp\to jj}\simeq\sigma^{8 \rm TeV}_{pp\to\tilde{\nu}_e}|\lambda_2^{1jk}|^2\cdot{\rm BR}_{d\bar{d}},\nonumber\\
\label{eq:snudijet8}
\end{align}
In addition, selectron decay $\tilde{e}_L^\pm \to u_j\bar{d}_k/\bar{u}_jd_k$ also contributes to dijet final states. Whose cross section can be well approximated by,
\begin{align}
\sigma(pp\to\tilde{e}_L^-\to \bar{u}_jd_k)\simeq\sigma^{8 \rm TeV}_{pp\to\tilde{e}_L^-}|\lambda_2^{1jk}|^2\cdot{\rm BR}_{\bar{u}d}, \nonumber\\
\sigma(pp\to\tilde{e}_L^+\to u_j\bar{d}_k)\simeq\sigma^{8 \rm TeV}_{pp\to\tilde{e}_L^+}|\lambda_2^{1jk}|^2\cdot{\rm BR}_{u\bar{d}}.
\label{eq:sedijet8}
\end{align}
Where ${\rm BR}_{\bar{u}d~(u\bar{d})}=\Gamma_{\bar{u}d~(u\bar{d})}/(\Gamma_{\bar{u}d~(u\bar{d})}+
\Gamma_{\chi l})$. $\Gamma_{\bar{u}d~(u\bar{d})}$ and $\Gamma_{\chi l}$ are obtained by replacing $m_{\tilde{\nu}_e}$ in Eq.~\ref{eq:dj} and \ref{eq:chi} to $m_{\tilde{e}_L^\mp}$, respectively. In Eq.~\ref{eq:sedijet8}, we have ignored the loop-induced decay modes $WZ$ and $W\gamma$ since they are high suppressed compared with tree-level ones. Notice that the mass splitting for the left-handed slpeton are determined by model independent relation $m^2_{\tilde{e}_L}-m^2_{\tilde{\nu}_e}=-\cos(2\beta)m^2_W$~\cite{Martin:1997ns}, we thus have $m_{\tilde{e}_L^\pm}=754$ GeV for $m_{\tilde{\nu}_e}=750$ GeV and $\tan\beta=10$. Which indicates contribution coming from $\tilde{e}_L^\pm$ must be taken into account since their masses are close to $\tilde{\nu}_e$. Similar to table~\ref{tab:pro_snu}, cross sections of $\tilde{e}_L^\pm$ production at LHC 8 TeV are listed in table~\ref{tab:pro_se}.

\begin{figure}[!htbp]
\begin{center}
~~
\includegraphics[width=0.88\linewidth]{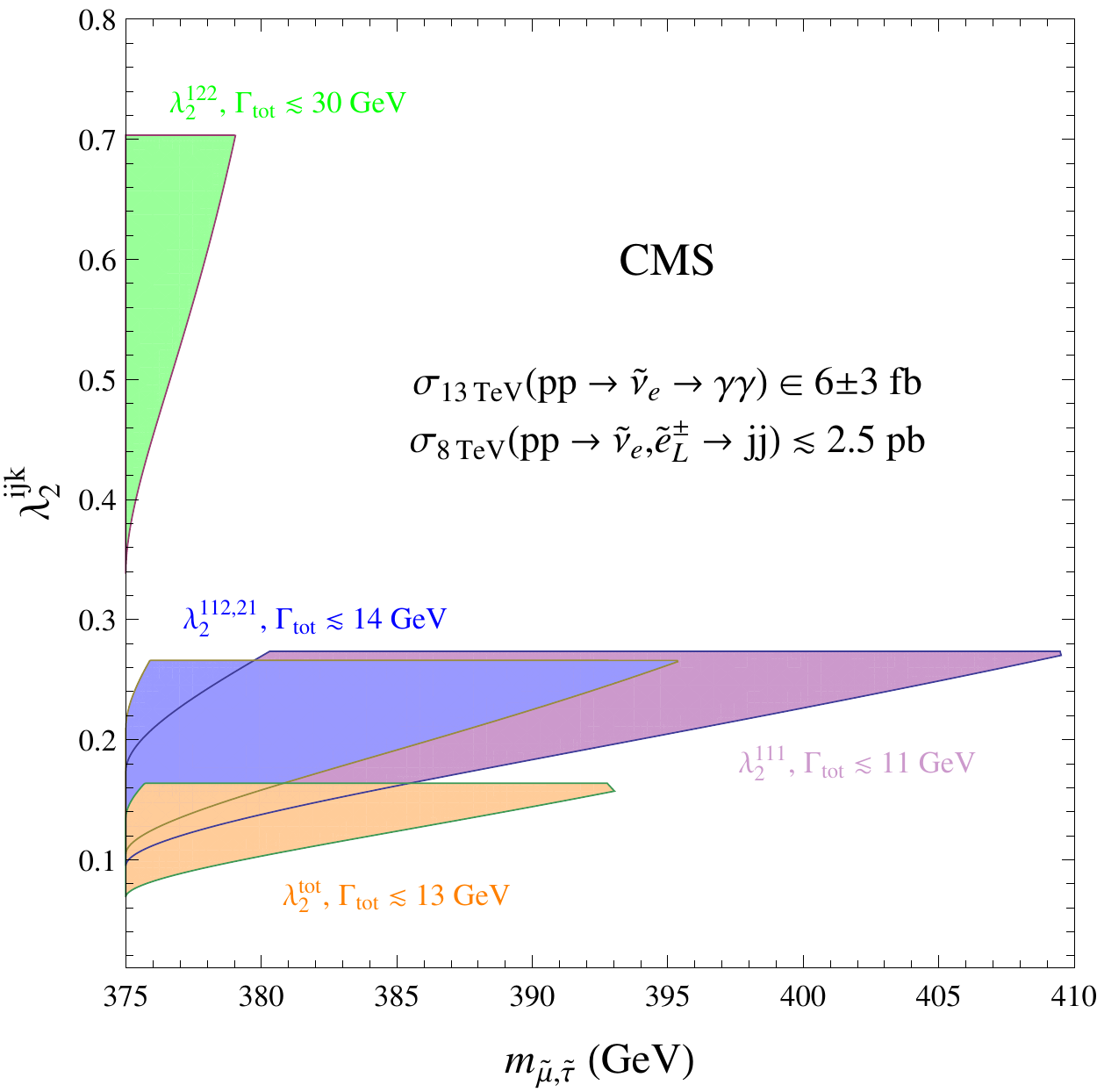}
\end{center}
\caption{The allowed region for the diphoton excess required by CMS observation on the $[m_{\tilde\mu,\tilde\tau},~\lambda_2^{ijk}]$ plane with only considering constraint from dijet constraint at 8 TeV LHC. Here the mass of electroweakinos  are fixed as $m_{{\tilde\chi}^{\pm},\tilde{\chi}^{0}}=350$ GeV. The regions with purple, green, blue and orange color correspond to contributions of $\lambda_2^{111}$, $\lambda_2^{122}$, $\lambda_2^{112,21}$ and $\lambda_2^{\rm tot}$, respectively. For comparison, the largest total decay width for each parameter region is also shown.
\label{fig:cms1}}
\end{figure}

\begin{figure}[!htbp]
\begin{center}
~~
\includegraphics[width=0.88\linewidth]{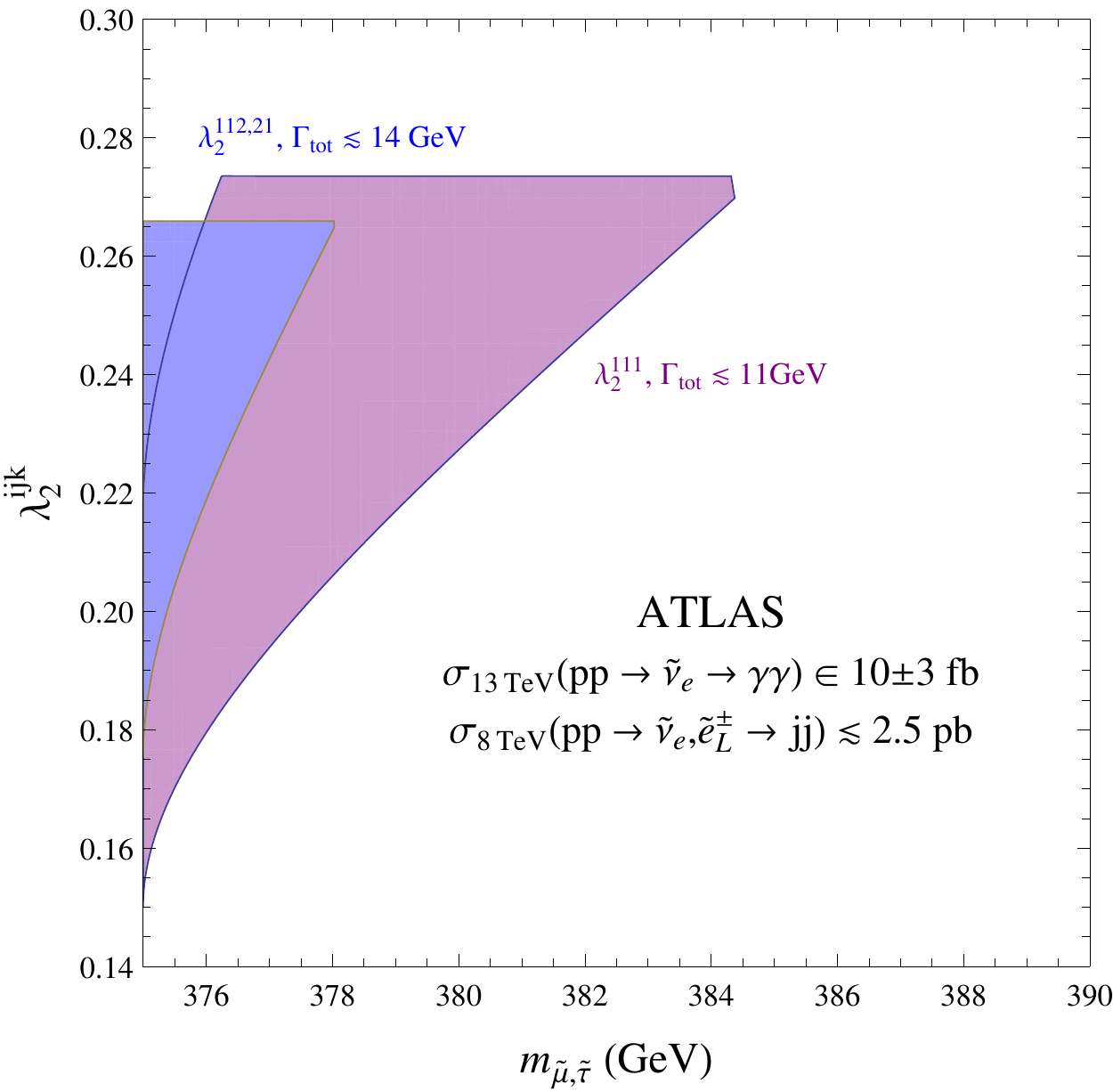}
\end{center}
\caption{Similar with figure~\ref{fig:cms1}, but for the ATLAS observation. Notice that in this case the parameter space for $\lambda_2^{122}$ and $\lambda_2^{\rm tot}$ are totally excluded thus do not presented here.
\label{fig:atlas}}
\end{figure}

We first investigate the case of light electroweakinos. We require that signal cross sections respectively satisfy the CMS
and ATLAS observations which are listed in Eq.~\ref{eq:sigma813}, with imposing the dijet upper bound $\sigma^{8 \rm TeV}_{pp\to jj}<2.5$ pb. In figure~\ref{fig:cms1} and~\ref{fig:atlas}, we respectively display the allowed regions for each independent operator in $[m_{\tilde\mu,\tilde\tau},~\lambda_2^{ijk}]$ plane for CMS and ATLAS observations. In addition, the largest total decay width can be reached in each parameter regions are also shown. There are some important results can be learned from these figures:
\begin{itemize}
  \item {All the independent couplings can fit the CMS observation. The upper bounds of allowed regions for different couplings $\lambda_2$ have inverse hierarchy with that of production rate in Eq.~\ref{eq:pro} since a smaller production cross section makes a larger upper bound of coupling. Furthermore, $\lambda_2^{111}$ and $\lambda_2^{112,21}$ give roughly comparable parameter space while $\lambda_2^{122}$ and $\lambda_2^{\rm tot}$ coupling hold relatively small regions. Which is due to the fact that $\lambda_2^{122}$ has the smallest production cross section, making it can only fit the signal rate in a narrow mass regions. On the other hand, $\lambda_2^{\rm tot}$ gives the largest senutrino production with the smallest diphoton branching ratio. As a result, it takes moderate diphoton signal rate. While the dijet constraint is most stringent for $\lambda_2^{\rm tot}$, thus significantly reduces its parameter space.}
  \item {For ATLAS observation, only $\lambda_2^{111}$ and $\lambda_2^{112,21}$ couplings are survived with tiny parameter space while $\lambda_2^{122}$ and $\lambda_2^{\rm tot}$ couplings are totally excluded. This can be explained as follows: ATLAS observation prefer large signal cross section, while senutrino production resulted from $\lambda_2^{122}$ coupling is too small to reach such signal rate. On the other hand, for $\lambda_2^{\rm tot}$ coupling, the reason is similar with the case of CMS. But for ATLAS case, required diphoton signal rate and dijet limit can never be balanced thus totally exclude $\lambda_2^{\rm tot}$ in whole parameter space.}
 \item {The largest total decay width corresponding to CMS (ATLAS) observation is about $30$ ($14$) GeV, which is provided by $\lambda_2^{122}$ ($\lambda_2^{112,21}$) coupling. Especially, best-fit width $\Gamma_{\rm tot}=45$ GeV suggested by ATLAS observation can never be satisfied in our parameter space.}
\end{itemize}

\begin{table}[hbtp]
\begin{tabular}{|c|c|c|c|c|}
\hline
Production cross section (pb)& $\lambda_2^{111}$ & $\lambda_2^{122}$ & $\lambda_2^{112,21}$  & $\lambda_2^{\rm tot}$   \\\hline
$\sigma^{13 \rm TeV}_{pp\to\tilde{\nu}_e}$ & $77.6$ & $9.6$ & $68.3$ & $155.4$ \\\hline
$\sigma^{8 \rm TeV}_{pp\to\tilde{\nu}_e}$ & $30$ & $2.3$ & $22.2$ & $54.6$ \\\hline
\end{tabular}
\caption{Production cross sections of $\tilde{\nu}_e$ at LHC 13 TeV and 8 TeV respectively correspond to couplings $\lambda_2^{111}$, $\lambda_2^{122}$, $\lambda_2^{112,21}$ and $\lambda_2^{\rm tot}$. Where the typical value is taken to be $|\lambda_2|=1$.}
\label{tab:pro_snu}
\end{table}

\begin{table}[hbtp]
\begin{tabular}{|c|c|c|c|c|}
\hline
Production cross section (pb)& $\lambda_2^{111}$ & $\lambda_2^{122}$ & $\lambda_2^{112,21}$  & $\lambda_2^{\rm tot}$   \\\hline
$\sigma^{8 \rm TeV}_{pp\to\tilde{e}_L^-}$  & $23.1$ & $1.4$ & $15.2$ & $45.9$ \\\hline
$\sigma^{8 \rm TeV}_{pp\to\tilde{e}_L^+}$  & $58.1$ & $3.2$ & $41.5$ & $144.6$ \\\hline
\end{tabular}
\caption{Similar with table~\ref
{tab:pro_snu}, but for $\tilde{e}_L^{\pm}$ at LHC 8 TeV.}
\label{tab:pro_se}
\end{table}

\begin{figure}[!htbp]
\begin{center}
~~
\includegraphics[width=0.88\linewidth]{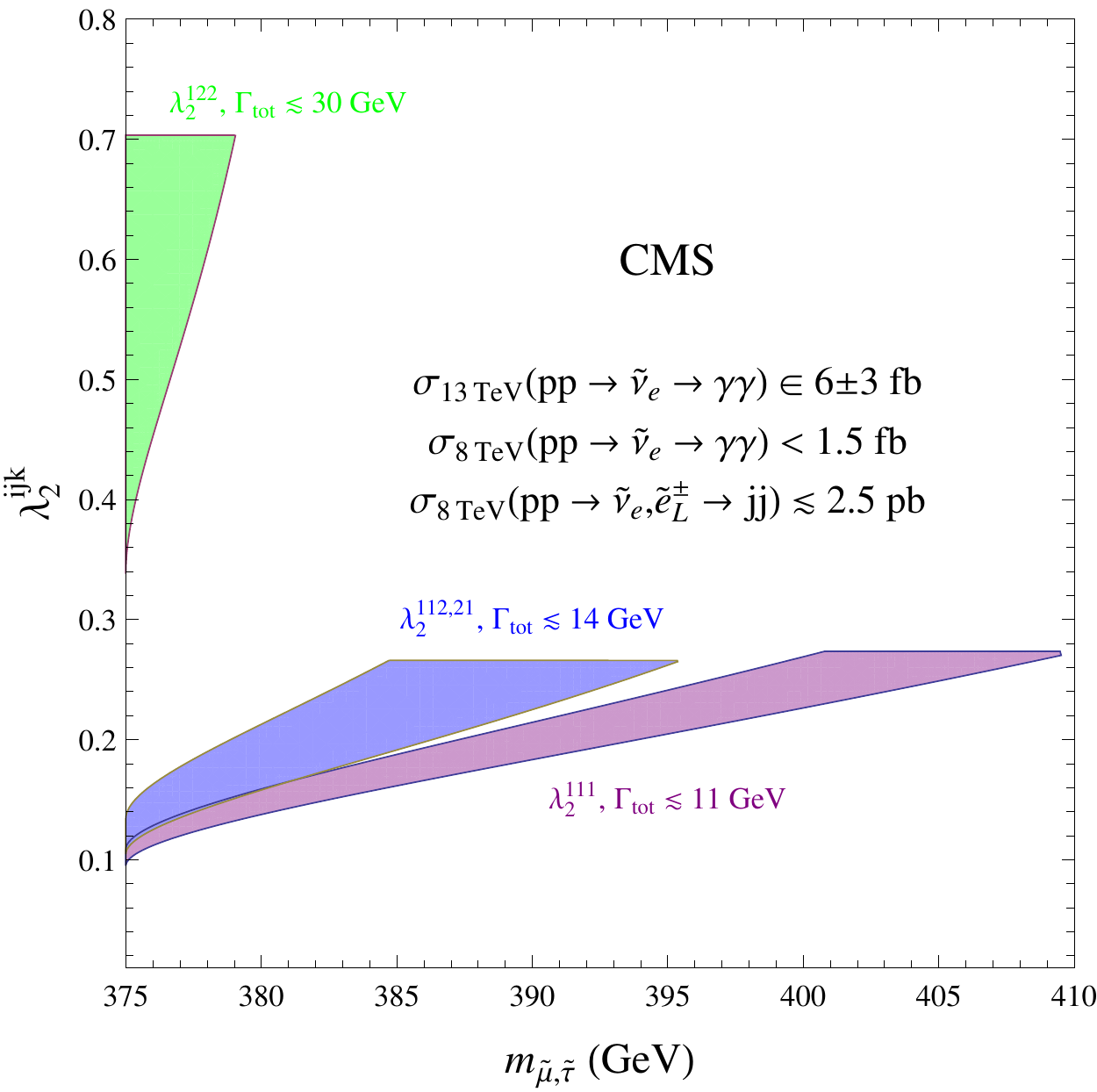}
\end{center}
\caption{The allowed region for the diphoton excess required by CMS observation on the $[m_{\tilde\mu,\tilde\tau},~\lambda_2^{ijk}]$ plane with considering both dijet and diphoton constraints at 8 TeV LHC.
\label{fig:cms2}}
\end{figure}

We further address constraint from diphoton channel. Since current upper limit for this channel is $\sigma^{8 \rm TeV}_{pp\to \gamma\gamma}<1.5$ fb, compatibility of diphoton excess between LHC $8$ TeV and $13$ TeV has to be examed. In figure~\ref{fig:cms2}, we present allowed parameter space in the same plane with imposing both dijet and diphoton constraints. where the diphoton signal rate at LHC 8 TeV is computed as
\begin{align}
\sigma^{8 \rm TeV}_{pp\to \gamma\gamma}\simeq\sigma^{8 \rm TeV}_{pp\to\tilde{\nu}_e}|\lambda_2^{1jk}|^2\cdot{\rm BR}_{\gamma\gamma}.
\end{align}
In this case, we find that the parameter space for ATLAS observation is entirely ruled out. Even for CMS result, allowed parameter regions for different couplings are also changed dramatically. To be specific, the parameter space is disappeared for $\lambda_2^{\rm tot}$ coupling; getting smaller for $\lambda_2^{111}$ and $\lambda_2^{112,21}$ couplings; while keeping almost unchanged for $\lambda_2^{122}$ coupling. To understand such behaviors, one noticed that the growth of production cross sections at LHC $13$ TeV and $8$ TeV are different for the first and second generation quarks. From table~\ref{tab:pro_snu}, we find that the increasing of cross sections for $\tilde{\nu}_e$ production are given by,
\begin{eqnarray}
\frac{\sigma^{13}_{pp\to\tilde{\nu}_e}}{\sigma^8_{pp\to\tilde{\nu}_e}}(\lambda_2^{111})&\simeq2.58&,\;
\frac{\sigma^{13}_{pp\to\tilde{\nu}_e}}{\sigma^8_{pp\to\tilde{\nu}_e}}(\lambda_2^{112,21})\simeq3.07,\nonumber\\
\frac{\sigma^{13}_{pp\to\tilde{\nu}_e}}{\sigma^8_{pp\to\tilde{\nu}_e}}(\lambda_2^{122})&\simeq4.17&,\;
\frac{\sigma^{13}_{pp\to\tilde{\nu}_e}}{\sigma^8_{pp\to\tilde{\nu}_e}}(\lambda_2^{\rm tot})\simeq2.85.
\label{eq:ratio}
\end{eqnarray}
Due to different PDF dependence, the growth ratio for second generation quarks are higher than that of first generation, leading to the couplings involved with second generation quarks ($\lambda_2^{122}$ and $\lambda_2^{112,21}$) have more significant increasing on signal rate. Meanwhile, the growth of observed diphoton signal rate from LHC 8 TeV to 13 TeV are estimated as
\begin{equation}
\frac{\sigma^{\rm 13 TeV}_{pp\to\gamma\gamma}}{\sigma^{\rm 8 TeV}_{pp\to\gamma\gamma}} \simeq
\left\{\begin{array}{lll}
\frac{(10\pm3)~\fb}{1.5~\fb}\sim4.7-8.7&\hbox{for ATLAS},\\
\frac{(6\pm3)~\fb}{1.5~\fb}\sim2-6&\hbox{for CMS}.
\end{array}\right.
\label{eq:ratio813}
\end{equation}
Based on Eq.~\ref{eq:ratio} and \ref{eq:ratio813}, it is clearly that none of couplings in our model can fit the ATLAS observation and simultaneously guarantee the compatibility of diphoton excess between LHC $8$ TeV and $13$ TeV. On the other hand, the situation is much better for CMS observation. In this case, couplings $\lambda_2^{122}$ and $\lambda_2^{112,21}$ are more advantageous since they are benefit from large growth ratio of second generation quarks, thus preserving the compatibility of diphoton excess as much as possible. While for $\lambda_2^{111}$ coupling which only involved with first generation quarks, whose increasing of cross section is slightly larger than the lower bound of growth of CMS diphoton signal rate, thus disfavored by compatibility of diphoton excess. Finally, coupling $\lambda_2^{\rm tot}$ is totally excluded due to stringent dijet bound although  possessing second smallest of increasing.

\begin{figure}[!htbp]
\begin{center}
~~
\includegraphics[width=0.88\linewidth]{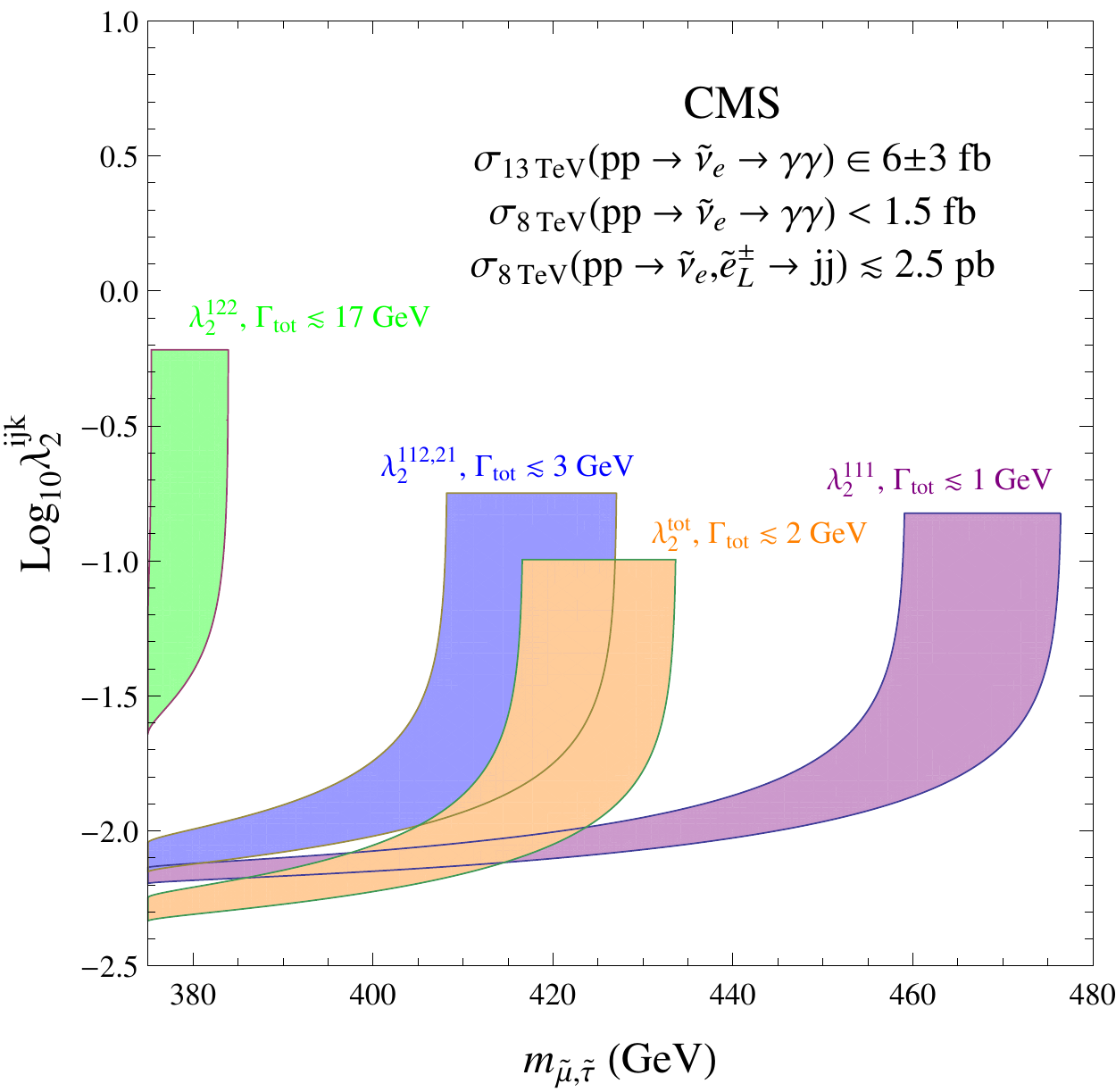}
\end{center}
\caption{The allowed region for the diphoton excess required by CMS observation on the $[m_{\tilde\mu,\tilde\tau},~\log\lambda_2^{ijk}]$ plane with considering constraints from both dijet and diphoton resonance searches. Here the mass of electroweakinos  are fixed as $m_{{\tilde\chi}^{\pm},\tilde{\chi}^{0}}=800$ GeV.
\label{fig:cms3}}
\end{figure}

Next, we briefly discuss the case of heavy electroweakinos. The remarkable difference is that the decay mode related to electroweakinos is kinematically forbidden in this case, leading to $\Gamma_{\chi l}=0$. Which clearly enhances the branching ratio of diphoton mode, while with the price of reducing total decay width. Similar with the case of light electroweakinos, the parameter space for ATLAS observation is also disappeared after imposing both dijet and diphoton constraints at LHC 8 TeV. We thus only present results on CMS observation in Fig.~\ref{fig:cms3}. The changing of parameter space are summarized as follows:
\begin{itemize}
  \item {the allowed mass regions for smuon/stau are extended up to about 480 GeV,}
  \item {the largest total decay width in each parameter regions are correspondingly reduced. For instance, $\Gamma_{\rm tot}$ respectively  drop to $17$ GeV and $1$ GeV for $\lambda_2^{122}$ and $\lambda_2^{111}$.}
\end{itemize}

Before ending up this section, we mention that the parameter space of $\lambda_2$ in Figs.~\ref{fig:cms2} and \ref{fig:cms3} should be also consistent with the exisiting low-energy constraints. The indirect bounds coming from various low energy experiments are collected in Ref.~\cite{Barbier:2004ez}, those most related to our model are listed blow~\footnote{We do not include limit coming from the neutrino masses and mixings since which relys on assumptions of the generation structure of $\lambda_2$~\cite{Barbier:2004ez}, which are not specified in our model. In addition, we adopt more recent result in Ref.~\cite{Allanach:2009xx} for limit from neutrinoless double beta decay.}:
\begin{eqnarray}
\lambda_2^{111} \ &\leq& \ 5\times 10^{-4}\left(\frac{m_{\tilde f}}{100~{\rm GeV}}\right)^2\left(\frac{m_{{\tilde g}/{\tilde \chi}}}{100~{\rm GeV}}\right)^\frac{1}{2}~[\beta\beta0\nu],\nonumber\\
\lambda_2^{11k} \ &\leq& \ 0.02\left(\frac{m_{\tilde d_{kR}}}{100~{\rm GeV}}\right)~[V_{ud}],\nonumber\\
\lambda_2^{12k} \ &\leq& \ 0.21\left(\frac{m_{\tilde d_{kR}}}{100~{\rm GeV}}\right)~[A_{FB}],\nonumber\\
\lambda_2^{1j1}\ &\leq& \ 0.03\left(\frac{m_{\tilde u_{jL}}}{100~{\rm GeV}}\right)~[Q_W(\rm Cs)].
\label{eq:low}
\end{eqnarray}
In above equation, $m_{\tilde f}$ and $m_{{\tilde g}/{\tilde \chi}}$ respectively denote sfermion, gluino/neutralino masses. The notations in the brackets stand for
the corresponding low-energy constraints which are illustrated as follows:
\begin{itemize}
  \item $\beta\beta0\nu$: neutrinoless double beta decay of nuclei, constraining by the lower limit on the half-life of $^{76}$Ge isotope~\cite{KlapdorKleingrothaus:2000sn}~\cite{Allanach:2009xx};
  \item $V_{ud}$: charged current universality in the quark sector, constraining by the experimental value of the CKM matrix element $V_{ud}$~\cite{Hagiwara:2002fs};
  \item $A_{FB}$: forward-backward asymmetries of fermion pair production process via $Z$ boson resonant $e^+e^-\to Z^0 \to f \bar{f}$, constraining through the experimental values of $A^f_{FB}$ measured by CERN LEP-I~\cite{Hagiwara:2002fs};
  \item $Q_W(\rm Cs)$: deviation of the weak charge $Q_W$ to its SM value, constraining by the parity violating transitions in $^{133}$Cs~\cite{Rosner:2001ck}~\cite{Ginges:2003qt}.
\end{itemize}
One can see that most of low-energy constraints rely on squark and gluino masses
which are decoupled in our spectrum, thus are harmless for the parameter space of $\lambda_2$. The only threatening constraint coming from the $\beta\beta0\nu$, which involved with charged slepton and neutralino masses thus imposing very stringent bound on $\lambda_2^{111}$. In the case of light electroweakinos with $(m_{\tilde{e}_L},~m_{\tilde{\chi}^0})=(754,~350)$ GeV, the first equation of Eq.~\ref{eq:low} implies $\lambda_2^{111}\leq 0.053$. As a consequence, the allowed regions for $\lambda_2^{111}$ in Figs.~\ref{fig:cms2} are totally excluded. On the other hand, for light electroweakinos with $(m_{\tilde{e}_L},~m_{\tilde{\chi}^0})=(754,~800)$ GeV, the corresponding limit is $\lambda_2^{111}\leq 0.08$. Which still allowing some parameter space for
$\lambda_2^{111}$ and $\lambda_2^{\rm tot}$.

{\bf Conclusion}--In summary, we have proposed an supersymmetric explanation of the diphoton excess within the framework of LRPV MSSM. Where the 750 GeV resonance is identified as sneutrino and produced by $q\bar{q}$ initial state. With introducing relatively large trilinear soft parameter $T^{1jj}_{\lambda_1}$, the branching ratio of diphoton mode received significantly enhancement compared with conventional MSSM. The important features and predictions of our model are summarized as follows:
\begin{itemize}
  \item In our model, the sneutrino resonance is produced via $q\bar{q}$ annihilation. Which is distinct from models produced by $gg$ initial state. The two production channel can be distinguished in principle by using angular distributions with sufficiently large statistics.
  \item With considering dijet and diphoton constraints at LHC 8 TeV, our model can successfully fit the CMS data in sizeable parameter regions. While for large signal cross section and a width of $45$ GeV suggested by ATLAS data, the parameter space is entirely ruled out. Fitting the excess strongly favors the mass of stau and smuon within the range of $375-410$ ($375-480$) GeV for the case of light (heavy) electroweakinos. Which is
      a strong prediction for mass spectrum of slepton sector.
  \item The predicted total decay width can reaches $30$ ($17$) GeV for light (heavy) electroweakinos.
  \item The low-energy constraint from neutrinoless double beta decay imposes severe bound on coupling $\lambda_2^{111}$. Combined with the compatibility of diphoton excess between LHC $8$ TeV and $13$ TeV, the couplings involved with second generation quarks have more advantages.
\end{itemize}

{\bf Note added}--
After a few days we submitted our paper to arXiv, Ref.~\cite{Allanach:2015ixl} appeared, which explains the diphoton excess using the similar scenario. The difference is that the authors consider degenerate electron or moun sneutrino as the resonance and producing only through first generation quarks. In addition, they do not discuss the compatibility of diphoton excess between LHC $8$ TeV and $13$ TeV.

{\bf Acknowledgements}--
This research was supported in part by the Natural Science Foundation of China
under grant numbers 11135003, 11275246, and 11475238 (TL).


\end{document}